\documentclass[aps,twocolumn,pra,superscriptaddress,amsmath,showpacs,tightenlines]{revtex4-1}
\usepackage{amssymb}
\usepackage{amsmath}
\usepackage{graphicx}
\usepackage{subfigure}
\usepackage{natbib}
\usepackage{epsfig}
\usepackage{amsfonts}
\usepackage{mathrsfs}
\usepackage{CJK}
\usepackage{xcolor}
\usepackage[toc,page,title,titletoc,header]{appendix}

\begin{document}

\title{Dissipation and decoherence induced by collective dephasing in
  coupled-qubit system with a common bath}

\author{Z. H. \surname{Wang}}%
\affiliation{Beijing Computational Science Research Center, Beijing
  100084, China}

\author{Y. J. \surname{Ji}}%
\affiliation{Beijing Computational Science Research Center, Beijing
  100084, China}

\author{Yong \surname{Li}}%
\affiliation{Beijing Computational Science Research Center, Beijing
  100084, China} \affiliation{Synergetic Innovation Center of Quantum
  Information and Quantum Physics, University of Science and
  Technology of China, Hefei 230026, China}

\author{D. L. \surname{Zhou}}%
\affiliation{Beijing National Laboratory for Condensed Matter Physics,
  Institute of Physics, Chinese Academy of Sciences, Beijing 100190,
  China}
\begin{abstract}
  The longitudinal coupling of a system to the bath usually induces the pure
  dephasing of the system. In this paper, we study the collective
  dephasing induced dissipation and decoherence in a coupled-qubit
  system with a common bath. It is shown that, compared with the case
  of the same system with independent baths, the interference between
  the dephasing processes of different qubits induced by the common
  bath significantly changes the dissipation of the system. For the
  system of two coupled qubits, the interference leads to a faster decoherence
  in the non-single-excitation subspaces and a slower dissipation (and
  decoherence) in the single-excitation subspace. For the system of multiple
  coupled qubits, we also find the slower dissipation in the
  single-excitation subspace and obtain the decay rates of the first
  excited states for different system sizes numerically. All our
  results on collective dephasing induced dissipation can be explained
  based on a simple model with Fermi's golden rule.
\end{abstract}


\maketitle

\section{Introduction}

The dynamics of quantum open systems~\cite{HP} has attracted more and
more attentions in the fields of quantum optics and quantum
information~\cite{VR,man}. The influence of the bath on a
single qubit is characterized by two times: the relaxation time $T_1$
and the decoherence time $T_2$ satisfying
$1/T_2=1/2T_1+1/T_{\varphi}$, where $T_1$ is determined by the
transverse coupling to the bath with the frequency near
resonance with that of the qubit while $T_{\varphi}$, named as `` pure
dephasing time" ~\cite{YM,hauss}, arises from the longitudinal
coupling to the bath with much lower frequencies.

In the system of superconducting qubits~\cite{You1}, the main
intrinsic element that limits the decoherence time results from the
low-frequency noise~\cite{You2,YM1,KK, SM}. Similarly, in some biology
systems such as the excitations in pigment-protein
complexes~\cite{MB,MM}, the phonon modes in the bath have much
lower frequencies than the transition frequency of the excitations. In
such kinds of systems, how to efficiently suppress the decoherence
arising from the low frequency noise is a central task.

Besides the single system embedded in the bath, the emergent
properties of many-body open system are also at the frontier of many
research fields. Due to the cooperative effect in coupled system
with subsystems interacting with a correlated~\cite{You3,AD} or
common~\cite{MB1, YH, LD, JP, ZH,JP1, LM, FB} bath, it usually
exhibits some exotic collective phenomena. However, in most of the
above literatures, the authors mainly discussed the entanglement induced by
many-body interaction and the coupling to the bath, instead of
the interference effect arising from the common or correlated
bath(s). Actually, the interference effect in atomic ensemble
leads to the well-known ``single-photon superradiance" ~\cite{Dicke}
in which the decay rate of the single-excitation state is proportional
to the number of the atoms involved due to the collective effect. A natural
question is what will the collective dephasing behave in the coupled
systems when the subsystems share a common bath?

In this paper, we study the collective dephasing of multiple
coupled qubits sharing a common bath.  Due to the coupling
between the neighbour qubits and the interference effect induced by
the common bath, the collective dephasing leads to the dissipation of
the excited states and decoherence from the viewpoint of energy
representation of the system. Based on our previous work about the
interference effect in multi-channel dissipation of the
system~\cite{ZH}, we continue to investigate the dynamics of two
coupled-qubit with a common bath in detail. The system under
consideration has an excitation conservation character and shows
dramatic difference in the non-single-excitation and single-excitation
subspaces. In the non-single-excitation subspaces, the dephasing of
the two qubits interferes with each other constructively, leading to a
faster decoherence but without dissipation. On the contrary, the
dephasing of the qubits interferes instructively in the
single-excitation subspace, and it shows slower dissipation and
decoherence processes. To demonstrate the underlying
mechanism, we reformulate the master equation to include the
interference effect of the bath, and compare the signs between
the self-dephasing terms and their interference terms to explain
the faster and slower decay of the
system. Moreover, we map the system of two coupled qubits into a single-qubit model.
It is mapped into a spin-boson model~\cite{legget}
in the single-excitation subspace and into a pure dephasing model
in the non-single-excitation subspaces. Then, the
dynamics of the system can be obtained intuitively without detailed
calculation.

Furthermore, we extend our discussions to the system of multiple
coupled qubits. Similar to the two-qubit system, we find that
the decay rates of the eigen-states are much smaller when the qubits
share a common bath than the situation with independent baths. We
obtain the decay rates of the first excited states in the
single-excitation subspace by Fermi's golden rule~\cite{MO,MH} for
three-qubit system and by numerical calculation for the system of
more qubits.

The rest of the paper is organized as follows. In Sec.~\ref{model}, we
present our model as multiple coupled qubits interacting longitudinally with
the common or independent bath(s). In Sec.~\ref{two}, we
investigate the collective dephasing induced dissipation and
decoherence for two coupled qubits sharing a common bath in detail, and
extend the discussions to the system composed of multiple
coupled qubits in Sec.~\ref{multi}. In Sec.~\ref{conclusion}, we give
some remarks and draw the conclusion.

\section{The model}
\label{model}

\begin{figure}
  \centering
  \includegraphics[width=8cm]{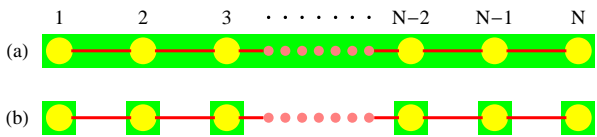}
  \caption{(Color online) The schematic diagram of the $N$ coupled
    identical qubits interacting longitudinally with a common bath (a) and
    with their independent baths (b). Here the qubits, the
    interaction between neighbor qubits, and the baths are
    represented by yellow balls, red links, and green rectangles
    respectively. }
  \label{scheme}
\end{figure}

As shown in Fig.~\ref{scheme}, the system under consideration is
composed of an align of $N$ identical coupled qubits, labelled by the
index $1,2,\cdots, N$, respectively, interacting with their common
[Fig.~\ref{scheme}(a)] or independent [Fig.~\ref{scheme}(b)]
bath(s).

The Hamiltonian of the system is
\begin{equation}
  H_S=\frac{\Omega}{2}\sum_{i=1}^{N} \sigma_{z}^{(i)}+\lambda\sum_{i=1}^{N-1}(\sigma_{-}^{(i)}\sigma_{+}^{(i+1)}+h.c.),
  \label{hs}
\end{equation}
where $\Omega$ is the energy level spacing between the ground state
$|g\rangle$ and excited state $|e\rangle$ of each qubit. $\sigma_z$,
$\sigma_-$ and $\sigma_+$ are the traditional Pauli operators for
two-level system. $\lambda$ is the coupling strength between arbitrary
two neighbor qubits.

Since any quantum system in nature cannot be absolutely isolated from
its surrounding bath, we must regard the quantum systems as
open systems~\cite{HP}. For a quantum system, when the
frequency of the mode in the bath is comparable to the
characteristic frequency of the system, it will induce the dissipation
and decoherence simultaneously. However, when the frequency of the
bath is much lower than that of the system, it contributes only
to the decoherence, which is named as pure dephasing. In the rest of
the paper, we will study the collective behavior of multiple coupled-qubit
system in such kinds of pure dephasing processes.

Firstly, we consider the situation where all the qubits share a common
bath as shown in Fig.~\ref{scheme}(a). The Hamiltonian of the whole
system then reads $H^{\rm{com}}=H_S+H_B^{\rm{com}}+H_I^{\rm{com}}$,
where $H_B^{\rm{com}}=\sum_j \omega_j b_j^{\dagger}b_j$ describes the
free terms of the bosonic modes in the bath. Here, $b_j$ is the
annihilation operator of the $j$th mode with frequency $\omega_j$. In
our consideration, it satisfies $\omega_j\ll \Omega$ for any $j$. The
interaction between the system and the bath is described by
\begin{equation}
  H_I^{\rm{com}}=\sum_i^{N}\sum_j \kappa_{i,j} \sigma_z^{(i)}(b_j^{\dagger}+b_j),
\end{equation}
where $\kappa_{i,j}$ is the coupling strength between the $i$th qubit
and the $j$th mode in the bath.

Secondly, for the case that each qubit interacts with its local
bath independently as shown in Fig.~\ref{scheme}(b), the
Hamiltonian is written as
$H^{\rm{ind}}=H_S+H_B^{\rm{ind}}+H_I^{\rm{ind}}$, with
$H_B^{\rm{ind}}=\sum_{i,j} \omega_{ij} c_{ij}^{\dagger}c_{ij}$ where
$c_{ij}$ is the annihilation operator of the $j$th mode in the
bath for the $i$th qubit and $\omega_{ij}$ ($\ll \Omega$) is
its frequency. The bath-system interaction is
\begin{equation}
  H_I^{\rm{ind}}=\sum_i^{N}\sum_j \kappa_{i,j} \sigma_z^{(i)}(c_{ij}^{\dagger}+c_{ij}).
\end{equation}

The complete information about the effect of the bath and the
coupling with the system is encapsulated in the spectral function
$J_i(\omega)$ which is defined by the expression~\cite{legget}
\begin{equation}
  J_i(\omega)\equiv\pi \sum_j \kappa_{i,j}^{2}\delta(\omega-\omega_j)
\end{equation}
for both the independent and common bath.

In what follows, we will consider the ohmic dissipation with the
spectrum functions being expressed as
\begin{equation}
  J_i(\omega)=\chi_i\omega\coth(\omega),
\end{equation}
where $\chi_i$ is the dissipation coefficient.

\section{Dynamics in two coupled-qubit system}
\label{two}

In this section, we will study the collective dephasing process of two
coupled qubits sharing a common bath, and analyze the quantum
interference effect in detail. As the central result of this paper, we
will show that the collective dephasing of the qubits leads to the
dissipation and decohernece of the system in the energy
representation.

\subsection{Master equation}
For the system of two coupled qubits, the Hamiltonian $H_S$ is reduced
to
\begin{equation}
  H_S=\frac{\Omega}{2}(\sigma_z^{(1)}+\sigma_z^{(2)})+\lambda(\sigma_-^{(1)}\sigma_+^{(2)}+h.c.).
\end{equation}
Correspondingly, the interaction between the qubits and their common
bath is written as
\begin{equation}
  H_I^{\rm{com}}=\sum_j (\kappa_{1j}\sigma_z^{(1)}+\kappa_{2j}\sigma_z^{(2)})(b_j^{\dagger}+b_j).
\end{equation}

We first diagonalize the Hamiltonian $H_S$, whose
eigen-states are
\begin{subequations}
  \begin{eqnarray}
    |\phi_1\rangle&=&|e;e\rangle,\\
    |\phi_2\rangle&=&|g;g\rangle,\\
    |\phi_{3}\rangle&=&\frac{1}{\sqrt{2}}(|e;g\rangle+|g;e\rangle),\\
    |\phi_{4}\rangle&=&\frac{1}{\sqrt{2}}(-|e;g\rangle+|g;e\rangle),
  \end{eqnarray}
\end{subequations}
and the corresponding eign-energies $E_i$ are
\begin{subequations}
  \begin{eqnarray}
    E_{1}&=&-E_2=\Omega,\\
    E_{3}&=&-E_4=\lambda,
  \end{eqnarray}
\end{subequations}
respectively.

\begin{figure}[t]
  \centering
  \includegraphics[scale=0.4]{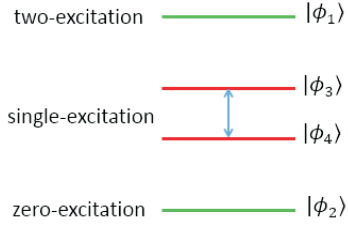}
  \caption{(Color online) The energy spectrum of $H_S$. The
    green lines represent the eigen-states in the non-single-excitation
  (i.e. two- and zero-excitation)
    subspaces while the red lines represent the eigen-states in the
    single-excitation subspace. The blue arrow implies the transition
    induced by $H_I^{\rm{com}}$. }
  \label{spectrum}
\end{figure}

In our system, the total excitations of the qubits are conserved with
$|\phi_3\rangle$ and $|\phi_4\rangle$ being in the single-excitation
subspace, while $|\phi_1\rangle$ and $|\phi_2\rangle$ being in the
subspaces with two and zero excitations respectively, which are
non-single-excitation subspaces. It is apparent that the interaction
between the qubits and the bath couples $|\phi_3\rangle$ and
$|\phi_4\rangle$, but not $|\phi_1\rangle$ and $|\phi_2\rangle$ as
shown in Fig.~\ref{spectrum}, where we illustrate the energy spectrum
of $H_S$.

Following the standard steps to derive the master equation~\cite{CC},
we obtain the master equation in the eigen-representation of $H_S$
as
\begin{equation}
  \frac{d\rho_{cd}}{dt}=-i\omega_{cd}\rho_{cd}+\sum_{k,l}\gamma^{cdkl}\rho_{kl}.
  \label{master}
\end{equation}
Here, $\rho_{cd}=\langle c|\rho|d\rangle$ ($c,d=\phi_i$, for
$i=1,2,3,4$) is the element of the reduced density matrix of the
coupled-qubit system, and $\omega_{cd}=E_c-E_d$ is the energy
difference between the states $|c\rangle$ and $|d\rangle$. In
Eq.~(\ref{master}), $\gamma^{cdkl}=\sum_{i=1}^{4} \gamma_i$ with

\begin{eqnarray}
  \gamma_{1}&=&-\delta_{dl}\sum_{n}\theta(\omega_{kn})[J_{1}(\omega_{kn})Z_{cn}^{(1)}Z_{nk}^{(1)}+J_{2}(\omega_{kn})Z_{cn}^{(2)}Z_{nk}^{(2)}\nonumber \\&&+\sqrt{J_{1}(\omega_{kn})J_{2}(\omega_{kn})}(Z_{cn}^{(2)}Z_{nk}^{(1)}+Z_{cn}^{(1)}Z_{nk}^{(2)})],\nonumber \\
  \gamma_{2}&=&\theta(\omega_{kc})[J_{1}(\omega_{kc})Z_{ck}^{(1)}Z_{ld}^{(1)}+J_{2}(\omega_{kc})Z_{ck}^{(2)}Z_{ld}^{(2)}\nonumber
  \\&&+\sqrt{J_{1}(\omega_{kc})J_{2}(\omega_{kc})}(Z_{ck}^{(2)}Z_{ld}^{(1)}+Z_{ck}^{(1)}Z_{ld}^{(2)})],\nonumber\\
  \gamma_{3}&=&-\delta_{ck}\sum_{n}\theta(\omega_{ln})[J_{1}(\omega_{ln})Z_{ln}^{(1)}Z_{nd}^{(1)}+J_{2}(\omega_{ln})Z_{ln}^{(2)}Z_{nd}^{(2)}\nonumber
  \\&&+\sqrt{J_{1}(\omega_{ln})J_{2}(\omega_{ln})}(Z_{ln}^{(2)}Z_{nd}^{(1)}+Z_{ln}^{(1)}Z_{nd}^{(2)})],\nonumber\\
  \gamma_{4}&=&\theta(\omega_{ld})[J_{1}(\omega_{ld})Z_{ck}^{(1)}Z_{ld}^{(1)}+J_{1}(\omega_{ld})Z_{ck}^{(2)}Z_{ld}^{(2)}\nonumber\\
  &&+\sqrt{J_{1}(\omega_{ld})J_{2}(\omega_{ld})}(Z_{ck}^{(2)}Z_{ld}^{(1)}+Z_{ck}^{(1)}Z_{ld}^{(2)})],
  \label{gamma}
\end{eqnarray}
where we have shorten $\sigma_z$ as $Z$ for the sack of compactness
and $A_{cd}=\langle c|A|d\rangle$ is the matrix element of the
operator $A$ in the energy representation of $H_S$. $\delta_{dl}$ and $\delta_{ck}$ are the usually used Dirac $\delta$ functions. The function $\theta(x)$ is
defined as
\begin{equation}
  \theta(x)=\begin{cases}
    1, & x\geq0\\
    0, & x<0
  \end{cases}.
\end{equation}
In Eq.~(\ref{gamma}), the terms proportional to
$\sqrt{J_1(\cdot)J_2(\cdot)}$ represent the contribution of the
quantum interference between the two dephasing channels.

\subsection{Collective dephasing induced dissipation}
Firstly, we confine ourselves in the non-single-excitation subspaces,
then the off-diagonal elements of the density matrix satisfy
\begin{equation}
  \frac{d}{dt}\rho_{12}=-2i\Omega \rho_{12}-\Gamma_{12}\rho_{12},
\end{equation}
whose solution is given by
\begin{equation}
  \rho_{12}(t)=\rho_{12}(0)\exp[(-2i\Omega-\Gamma_{12})t].
  \label{rhoo12}
\end{equation}
Here the decay rate of $\rho_{12}$ is
\begin{eqnarray}
  \Gamma_{12}&=&4[\sqrt{J_1(0)}+\sqrt{J_2(0)}]^2\nonumber \\
  &=&4[J_1(0)+J_2(0)+2\sqrt{J_1(0)J_2(0)}].
  \label{rho12}
\end{eqnarray}
Besides, the diagonal elements of the density matrix satisfy
$d\rho_{11}/dt=d\rho_{22}/dt=0$, that is, $\rho_{11}$ and $\rho_{22}$
do not decay during the time evolution.

It shows in Eq.~(\ref{rhoo12}) that the dephasing of the two qubits
actually leads to the decoherence of the system in the energy
representation. Attentions should be paid that the last term in the
second line of Eq.~(\ref{rho12}) comes from the quantum interference
effect between the two dephasing channels arising from the common
bath. When the two qubits interact with their local baths, this
term will disappear, and the decay rate becomes
$\Gamma_{12}^{\prime}=4[J_1(0)+J_2(0)]$, which is smaller than
$\Gamma_{12}$. Therefore, the common bath enhances the
decoherence of the system in the energy representation.

The mechanism underlying the faster decoherence can be observed in
Eq.~(\ref{gamma}). Let us explain the quantum interference effect from
the expressions of $\gamma_i$. In the non-single-excitation
subspaces, the terms resulted from the interference
of the dephasing of the two qubits have the same sign with
their self-dephasing term, that is \begin{equation}Z_{ii}^{(1)}Z_{ii}^{(2)}=Z_{ii}^{(2)}Z_{ii}^{(1)}=Z_{ii}^{(1)}Z_{ii}^{(1)}=Z_{ii}^{(2)}Z_{ii}^{(2)}=1 \label{construct}\end{equation}
for $i=1,2$. Therefore, the dephasing processes of two qubits
constructively interfere with each other, leading to a faster
decoherence.

Next, let us move to the single-excitation subspace. Then the elements
of the reduced density matrix satisfy
\begin{eqnarray}
  \frac{d}{dt}\rho_{33}&=&-2\Gamma_{33}\rho_{33},\\
  \frac{d}{dt}\rho_{34}&=&-2i\lambda\rho_{34}+\Gamma_{34}(\rho_{43}-\rho_{34}),
\end{eqnarray}
where the decay rate $\Gamma_{33}$ and decoherence rate $\Gamma_{34}$
are
\begin{eqnarray}
  \Gamma_{33}&=&\Gamma_{34}=[\sqrt{J_1(\omega_{34})}-\sqrt{J_2(\omega_{34})}]^2\nonumber \\
  &=&J_1(\omega_{34})+J_2(\omega_{34})-2\sqrt{J_1(\omega_{34})J_2(\omega_{34})}.
  \label{rho34}
\end{eqnarray}
\begin{figure}[tbp]
  \begin{centering}
    \includegraphics[width=8cm]{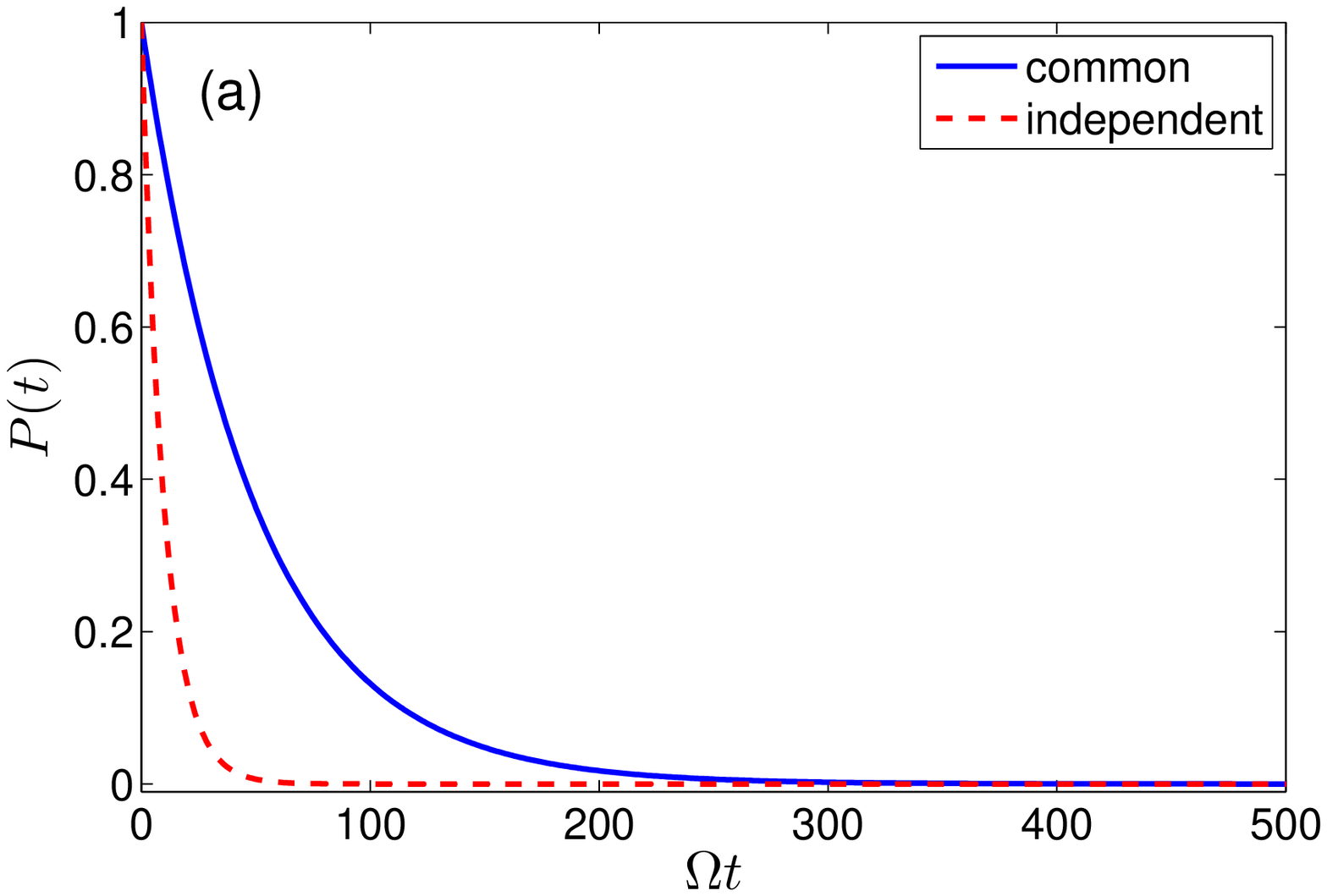}
    \par\end{centering}
  \begin{centering}
    \includegraphics[width=8cm]{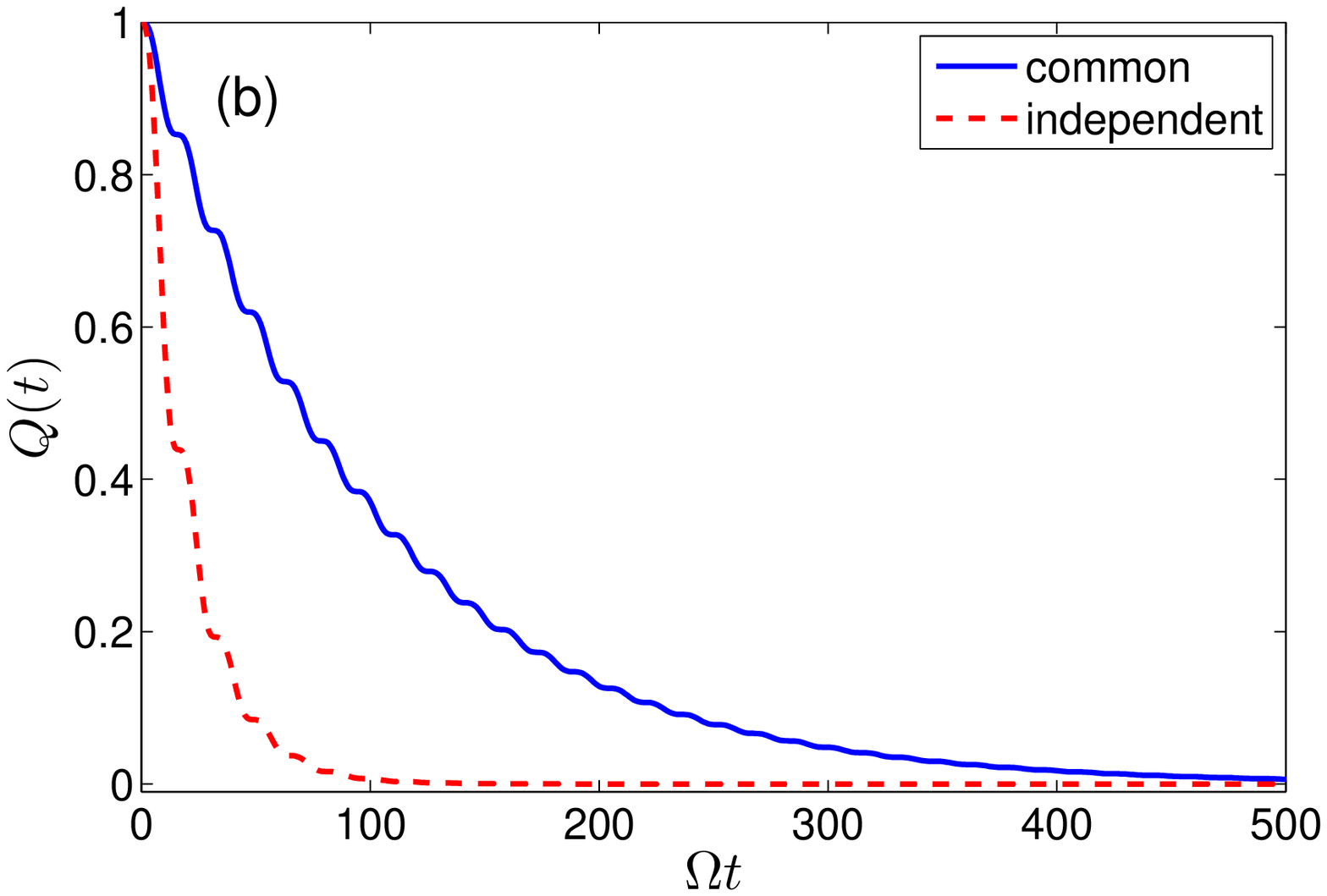}
    \par\end{centering}
  \caption{(Color online) The time evolution of (a) $P(t)$ and
    (b) $Q(t)$. The parameters are set as $\lambda=0.1,
    \chi_{1}=0.04$, and $\chi_2=0.01$ in units of $\Omega=1$. }
  \label{sub}
\end{figure}

The last term in Eq.~(\ref{rho34}), which comes from the interference
of dephasing between the two qubits, slows down the dissipation and decoherence of
the whole system. Especially, when the spectra satisfy
$J_1(\omega_{34})=J_2(\omega_{34})$, the single-excitation subspace is
a decoherence free subspace~\cite{PZ,DA,JK}, in which there is neither
dissipation nor decoherence.

To clearly show the dissipation and decoherence of the system in
the single-excitation subspace, we define the variables
\begin{align}
  P(t):=\frac{|\rho_{33}(t)|}{|\rho_{33}(0)|}, &&
  Q(t):=\frac{|\rho_{34}(t)|}{|\rho_{34}(0)|},
\end{align}
and plot the time evolution of $P(t)$ and $Q(t)$ in Fig.~\ref{sub}
with the assumption that the system is initially prepared in the state
$|e;g\rangle$. For comparison, we consider both the cases in which the
two coupled qubits interact with a common (blue solid line) and two independent baths
(red dashed line).
It is shown in Fig.~\ref{sub} that, when the two qubits share a
common bath, the decays of $\rho_{33}$ and
$\rho_{34}$ are much slower than the case with independent bathes due
to the instructive interference. The instructive interference comes
from the fact that
\begin{equation}
  Z_{ij}^{(1)}Z_{ij}^{(2)}=Z_{ij}^{(2)}Z_{ij}^{(1)}=-Z_{ij}^{(1)}Z_{ij}^{(1)}=-Z_{ij}^{(2)}Z_{ij}^{(2)}=-1
  \label{instruct}
\end{equation}
for $i,j=3,4$. It is just opposite with Eq.~(\ref{construct}) that
the interference term
$Z_{ij}^{(1)}Z_{ij}^{(2)}$ and $Z_{ij}^{(2)}Z_{ij}^{(1)}$ have
opposite signs with the self-dephasing terms $Z_{ij}^{(1)}Z_{ij}^{(1)}$ and
$Z_{ij}^{(2)}Z_{ij}^{(2)}$. Therefore, the interference effect leads to a slower dissipation and decoherence in the single-excitation subspace.

\subsection{Mapping to a single qubit}

For the situation of two coupled qubits sharing a common
bath which couples to the qubits longitudinally, it can be mapped to the model of a single qubit exposing in the
bath.

For the system of two coupled qubits, the interaction between the qubits and the common bath is written as
\begin{eqnarray}
  H_I^{\rm{com}}&=&\sum_j\frac{\kappa_{1j}+\kappa_{2j}}{2} (\sigma_z^{(1)}+\sigma_z^{(2)})(b_j+b_j^{\dagger})\nonumber \\&&+\sum_j\frac{\kappa_{1j}-\kappa_{2j}}{2} (\sigma_z^{(1)}-\sigma_z^{(2)})(b_j+b_j^{\dagger}).
\end{eqnarray}

In the single-excitation subspace, it is obvious that
$\sigma_z^{(1)}+\sigma_z^{(2)}=0$, then the corresponding
interaction Hamiltonian becomes
\begin{eqnarray}
  H_{I,s}^{\rm{com}}&=&\sum_j\frac{\kappa_{1j}-\kappa_{2j}}{2} (\sigma_z^{(1)}-\sigma_z^{(2)})(b_j+b_j^{\dagger})\nonumber \\
  &=&-(|\phi_3\rangle\langle\phi_4|+|\phi_4\rangle\langle\phi_3|)\sum_j(\kappa_{1j}-\kappa_{2j})(b_j+b_j^{\dagger}),\nonumber \\
\end{eqnarray}
where the second subscribe ``$s$" implies the single-excitation subspace.
We further define the collective operators
\begin{eqnarray}
  J_z&=&|\phi_3\rangle\langle\phi_3|-|\phi_4\rangle\langle\phi_4|,\\
  J_x&=&|\phi_3\rangle\langle\phi_4|+|\phi_4\rangle\langle\phi_3|.
\end{eqnarray}

Then the total Hamiltonian in the single-excitation subspace becomes
\begin{equation}
  H_{s}=\lambda J_z+\sum_j\omega_j b_j^{\dagger}b_j-J_x\sum_j g_j(b_j+b_j^{\dagger})
\end{equation}
which is a standard spin-boson model~\cite{legget}, with the coupling
strength between the ``spin" and the bosonic bath being
$g_j=\kappa_{1j}-\kappa_{2j}$. Then both the decay and decoherence rates
of the ``spin" in the energy representation are proportional to
$\sum_j(g_j)^{2}$. This result coincides with the results from the
master equation [see Eq.~(\ref{rho34})].

In the non-single-excitation subspaces, it satisfies that
$\sigma_z^{(1)}-\sigma_z^{(2)}=0$, and the interaction Hamiltonian becomes
\begin{eqnarray}
  H_{I,ns}^{\rm{com}}&=&\sum_j\frac{\kappa_{1j}+\kappa_{2j}}{2}(\sigma_z^{(1)}+\sigma_z^{(2)})(b_j^{\dagger}+b_j)\nonumber\\
  &=&(|\phi_1\rangle\langle\phi_1|-|\phi_2\rangle\langle\phi_2|)\sum_j(\kappa_{1j}+\kappa_{2j})(b_j^{\dagger}+b_j).\nonumber \\
\end{eqnarray}
where the second subscribe ``$ns$" implies the non-single-excitation subspaces.

Similar to the case in single-excitation subspace, we can define the
collective operator
\begin{equation}
  \mathcal{J}_z=|\phi_1\rangle\langle\phi_1|-|\phi_2\rangle\langle\phi_2|,
\end{equation}
then the total Hamiltonian in the non-single-excitation subspaces yields
\begin{equation}
  H_{ns}=\Omega\mathcal{J}_z+\sum_j\omega_jb_j^{\dagger}b_j
  +\mathcal{J}_z\sum_j\xi_j(b_j^{\dagger}+b_j)
\end{equation}
which is a pure dephasing Hamiltonian for a ``spin" with the coupling
strength between the spin and the bosonic bath being
$\xi_j=\kappa_{1j}+\kappa_{2j}$. Therefore, in the energy
representation, it experiences only the decoherence but without
dissipation and the decoherence rate is
$\gamma_\varphi\propto\sum_j(\xi_j)^2$. This fact also coincides with
the results from the master equation [see Eq.~(\ref{rho12})].

\section{Dynamics in multiple coupled-qubit system}
\label{multi}

In Sec.~\ref{two}, we have studied the collective dephasing of two
coupled qubits with a common bath. Now, we generalize it to the system
of $N$ ($>2$) coupled qubits.

For the system of $N$ qubits, the eigen-energies of the Hamiltonian $H_S$
in Eq.~(\ref{hs}) in the single-excitation subspace are
\begin{equation}
  E_n=\Omega(1-\frac{N}{2})+2\lambda\cos(\frac{n\pi}{N+1}),
\end{equation}
where $n=1,2,\cdots,N$ is taken as integers and the corresponding
eigen-states are
\begin{equation}
  |\psi_n\rangle=\sqrt{\frac{2}{N+1}}\sum_{j=1}^{N}\sin(\frac{nj\pi}{N+1})\sigma_+^{(j)}|G\rangle,
\end{equation}
where $|G\rangle$ represents that all the qubits are in their ground
states.

We further write the interaction between the qubits and the common
bath as
\begin{equation}
  H_I^{\rm{com}}=\sum_{\alpha=1}^{N} Z^{(\alpha)}\otimes R^{(\alpha)},
\end{equation}
where $R^{(\alpha)}=\sum_j \kappa_{\alpha,j}(b_j^{\dagger}+b_j)$. As
before, we have written $\sigma_z^{(\alpha)}$ as $Z^{(\alpha)}$ for compactness. The
Lindblad master equation under Born-Markov and secular approximations
is expressed as~\cite{HP}
\begin{equation}
  \frac{d}{dt}\rho=-i[H_s,\rho]+D[\rho]\label{masterm}
\end{equation}
with
\begin{eqnarray}
  D[\rho]&=&\sum_{\omega,\alpha,\beta}\gamma_{\alpha\beta}(\omega)[\tilde{Z}_\beta(\omega)\rho\tilde{Z}^\dagger_\alpha(\omega)-
  \frac{1}{2}\tilde{Z}^\dagger_\alpha(\omega)\tilde{Z}_\beta(\omega)\rho\nonumber \\&-&
  \frac{1}{2}\rho\tilde{Z}^\dagger_\alpha(\omega)\tilde{Z}_\beta(\omega)],
  \label{lindblad}
\end{eqnarray}
where
\begin{equation}
  \gamma_{\alpha\beta}(\omega)=\begin{cases}
    \sqrt{J_{\alpha}(\omega)J_{\beta}(\omega)}, & \omega\geq0\\
    0, & \omega<0
  \end{cases},
\end{equation}
and
\begin{equation}
  \tilde{Z}_\alpha(\omega)=\sum_{E_n-E_m=\omega}\langle \psi_{m}|Z^{(\alpha)}|\psi_{n}\rangle|\psi_m\rangle\langle \psi_n|.
\end{equation}

In the summation of Eq.~(\ref{lindblad}), the terms with
$\beta=\alpha$ come from the dephasing of the individual qubits, which
are always being for either the case that the qubits share a common
bath or interact with their local baths independently.
However, the terms for $\beta\neq\alpha$, which come from the
interference of dephasing processes for different qubits, exist only
when the qubits share a common bath.

\begin{figure}[tbp]
  \centering
  \includegraphics[width=8cm]{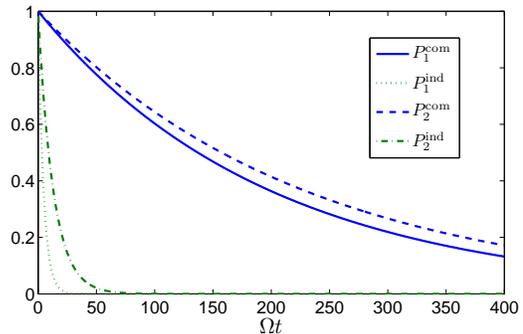}
  \caption{(Color online) The probability $P_1$ and $P_2$ as functions
    of the evolution time $t$ for the system of three coupled qubits. The coupling strength between the neighbour qubits is
    set as $\lambda=0.2$, and the dissipation coefficients are chosen as $\
    \chi_{i}=0.1\sin(i\pi/2N)$ ($i=1\cdots N$). All the parameters are in units of
    $\Omega=1$. }
  \label{three}
\end{figure}

Now, let us take the case of $N=3$ as an example to show how the
interference between the dephasing processes of different qubits 
induced by the common bath affects the dissipation of the eigen-states 
of the system. To this
end, we prepare the system in the state $|\psi_i\rangle$ ($i=1,2$)
initially and calculate the probability $P_i$ for the system still in
the initial state respectively. By solving the master equation [Eq.~(\ref{masterm})]
numerically, we plot $P_i$ as functions of the evolution time in Fig.~\ref{three}. For
comparison, we also plot the results when the different qubits
interact with their local baths independently. In Fig.~\ref{three}, we use the notations
 $P_i^{\rm{com}}$ and $P_i^{\rm{ind}}$ to represent the situation with
 common and independent baths, respectively.  It is clearly shown
that, the common bath induced interference slows down the dissipation
of the system. To study the interference effect in detail, we apply Fermi's
golden rule~\cite{MO,MH} to calculate the decay rates of $P_1$ and
$P_2$ which are denoted by $\gamma_1$ and $\gamma_2$ respectively. The
results are given by $\gamma_{i} =
\gamma_{i}^{\mathrm{ind}}+\gamma_{i}^{\mathrm{int}}$ $(i=1,2)$ where
\begin{subequations}
\begin{eqnarray}
  \gamma_1^{\mathrm{ind}}& = & \frac{1}{2}J_{1}(\omega_{12})+\frac{1}{4}J_{1}(\omega_{13})\nonumber \\&&+J_{2}(\omega_{13})+\frac{1}{2}J_{3}(\omega_{12})+\frac{1}{4}J_{3}(\omega_{13}), \\ \gamma_2^{\mathrm{ind}} & =&\frac{1}{2}J_{1}(\omega_{23})+\frac{1}{2}J_{3}(\omega_{23}),\\
  \gamma_{1}^{\mathrm{int}} & = & -\sqrt{J_{1}(\omega_{12})J_{3}(\omega_{12})}-\sqrt{J_{1}(\omega_{13})J_{2}(\omega_{13})}\nonumber \\&&+\frac{1}{2}\sqrt{J_{1}(\omega_{13})J_{3}(\omega_{13})}-\sqrt{J_{2}(\omega_{13})J_{3}(\omega_{13})},\nonumber \\ \\
  \gamma_{2}^{\mathrm{int}} & = & -\sqrt{J_{1}(\omega_{23})J_{3}(\omega_{23}).}
\end{eqnarray}
\end{subequations}
In the above equations, the terms $\gamma_i^{\rm{ind}}$ are the
decay rates when the different qubits couple to their local baths independently, and
the terms $\gamma_i^{\rm{int}}$ represent the contribution of the interference arising from the common bath.  Obviously, the dephasing of each
qubit interferes with each other, leading to a slower dissipation of
the system in the single-excitation subspace. We have numerically
checked that the decay rates given by Fermi's golden rule coincide
with the results from the master equation very well.
\begin{figure}[tbp]
  \includegraphics[scale=0.3]{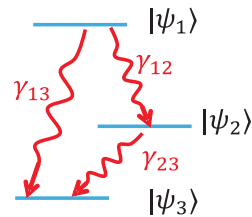}
  \caption{(Color online) Schematic illustration of decay of the
    excited states for three coupled-qubit system.}
  \label{decay}
\end{figure}
\begin{figure}[tbp]
  \centering
  \includegraphics[width=8cm]{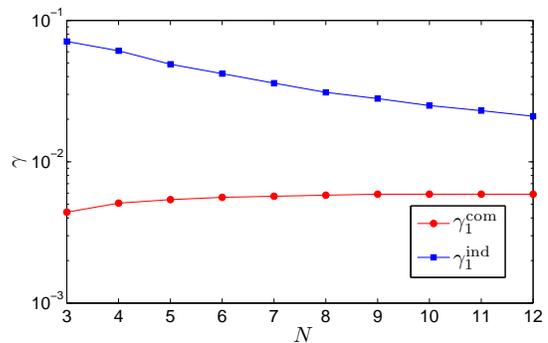}
  \caption{(Color online) The decay rates of the first excited state
    in the single-excitation subspace as a function of the number of
    qubits $N$. The parameters are the same as those in
    Fig.~\ref{three}. }
  \label{rate}
\end{figure}

In Fig.~\ref{decay}, we demonstrate the decay of the excited states
more intuitively. Due to the interaction between the bath and the
system, the excited state $|\psi_1\rangle$ will decay to the states
$|\psi_2\rangle$ and $|\psi_3\rangle$ simultaneously with the decay
rates $\gamma_{12}$ and $\gamma_{13}$ respectively. Meanwhile,
$|\psi_2\rangle$ will decay to $|\psi_3\rangle$ with the decay rate
$\gamma_{23}$. The decay rates $\gamma_{12},\gamma_{13}$, and
$\gamma_{23}$ can also be given approximately by Fermi's golden rule
for both independent and common bath. Then, the probabilities $P_i$
satisfy the equations
\begin{eqnarray}
  \frac{d}{dt}P_1&=&-(\gamma_{12}+\gamma_{13})P_1,\\
  \frac{d}{dt}P_2&=&-\gamma_{23}P_2+\gamma_{12}P_1.
\end{eqnarray}
From the above equations, we can observe that
$\gamma_1=\gamma_{12}+\gamma_{13}$ and $\gamma_2=\gamma_{23}$. That
is, $P_i$ $(i=1,2)$ exhibits an exponential decay when the system is
initially prepared in the state $|\psi_i\rangle$. It agrees with the
numerical predictions as shown before.

We can also extend our discussions to the system of $N$ ($>3$)
coupled qubits. In such a system, we calculate the decay rates of the
first excited states in the single-excitation subspace numerically. As
shown in Fig.~\ref{rate}, the decay rates are much lower when the
different qubits share a common bath (denoted by $\gamma_1^{\rm{com}}$ and represented by the red line with solid circles) than those with independent baths (denoted by $\gamma_1^{\rm{ind}}$ and represented by the blue line with solid rectangles). Furthermore, the decay rates decrease with the increase of the
number of the qubits $N$ when they interact with their local baths
independently. In contrast, it tends to be a constant as the increase
of $N$ for the case of common bath due to the complicated interference
processes among different qubits during the time evolution.

\section{Remarks and Conclusions}
\label{conclusion}

In experiments, our theoretical model can be realized in
superconducting qubits array system. In such kind of systems, the
neighbored qubits couple to each other via capabilities and the
dephasing of the qubits mainly comes from the $1/f$ noise, which is
induced by the charge or flux fluctuations~\cite{You2}.

In conclusion, we have investigated the collective dephasing behavior
in the system of coupled qubits interacting with the common and
independent baths. In the situation of the common bath, we show
that the collective dephasing will induce the dissipation and
decoherence of the system in the energy representation. On one hand,
we studied the system of two coupled qubit, which respectively shows
the slower dissipation (together with decoherence) in the
single-excitation subspace and faster dissipation (but without decoherence) in the
non-single-excitation subspaces. The exotic phenomena can be explained
from the viewpoint of quantum instructive and constructive
interference between the dephasing of the two qubits. On the other
hand, we studied the system composed of multiple qubits, and show how
the interference effect slows the dissipation of the system in the single-excitation subspace. Both for the system of two and multiple coupled qubits, our results can be 
understood from the viewpoint of Fermi's golden rule in the energy representation.

\begin{acknowledgments}
  We thank the helpful discussions from S. W. Li. This work is supported
  by NSF of China under Grants Nos.11121403 and 11174027. DLZ is supported by
  NSF of China under Grant No.11175247, and NKBRSF of China under
  Grants Nos.2012CB922104 and 2014CB921202.
\end{acknowledgments}

\end{document}